\documentclass[reprint,floatfix,aps,prd,showpacs,footinbib,amsmath,amssymb,amsfonts,superscriptaddress]{revtex4-1}

\usepackage{bm}
\usepackage{graphicx}
\usepackage{color}
\usepackage{dsfont}
\usepackage[normalem]{ulem}
\usepackage{hyperref}
\usepackage{mathrsfs}
\usepackage{float}
\usepackage{varioref}
\usepackage[active]{srcltx}
\expandafter\ifx\csname package@font\endcsname\relax\else
\expandafter\expandafter
\expandafter\usepackage
\expandafter\expandafter
\expandafter{\csname package@font\endcsname}%
\fi
\hypersetup{
colorlinks=true,       
linkcolor=red,          
citecolor=blue,        
}

\usepackage[section]{placeins}

\usepackage{fancybox}
\labelformat{equation}{Eq.(#1)} 
\labelformat{figure}{Fig.~#1} 
\labelformat{subfigure}{Fig.~\thefigure#1} 
\labelformat{table}{Tab.~#1} 
\newcommand{\be}{\begin{equation}}
\newcommand{\ee}{\end{equation}}
\newcommand{\bea}{\begin{eqnarray}}
\newcommand{\eea}{\end{eqnarray}}
\newcommand{\nn}{\nonumber}

\begin{document}


\title{Generalized thermalization for integrable system under quantum quench}

\author{Sushruth Muralidharan} \email{sm393@buffalo.edu}
\affiliation{Department of Physics, University at Buffalo, The State University of New York, Buffalo 14260, USA}
\author{Kinjalk Lochan}%
\email{kinjalk@iisermohali.ac.in}
\affiliation{Department of Physical Sciences, IISER Mohali, Manauli 140306, India}
\author{S. Shankaranarayanan}%
\email{shanki@phy.iitb.ac.in}
\affiliation{Department of Physics, Indian Institute of Technology Bombay, Mumbai 400076, India}

\begin{abstract}
 We investigate equilibration and {\it generalized thermalization} of the quantum Harmonic chain under
local quantum quench. The quench action we consider is connecting two disjoint harmonic chains
of different sizes and the system jumps between two integrable settings. We verify the validity of the
Generalized Gibbs Ensemble description for this \emph{infinite dimensional Hilbert space} system and 
also identify {\it equilibration between the
subsystems} as in classical systems. Using Bogoliubov transformations, we show that the eigenstates
of the system prior to the quench evolve towards the Gibbs Generalized Ensemble description.
Eigenstates that are more delocalized (in the sense of inverse participation ratio) prior to the quench,
tend to equilibrate more rapidly. Further, through the phase space properties of a Generalized Gibbs
Ensemble and the strength of stimulated emission, we identify the necessary criterion on the initial
states for such relaxation at late times and also find out the states which would potentially not be described by  the Gibbs Generalized Ensemble description.
\end{abstract}

\maketitle

\section{\label{sec:intro}INTRODUCTION}

Thermalization of isolated quantum systems is apparently
different from their classical counterparts \cite{CFLV}. In a
classical system, the constituent degrees of freedom interact and trade
off \emph{charges} to settle into a thermal equilibration configuration
\cite{gogolin,gge3,Polkovnikov}. Within the framework of unitary quantum theory, a system in a pure state will never evolve to a \emph{thermal state}. Instead, it has been suggested that {\it thermalization} of quantum systems needs to be understood through observables rather than the states themselves \cite{ergo}. To go about this, we need to separately quantify {\it equilibration} and {\it thermalization}. An observable is said to \textit{equilibrate} to a particular expectation value, within a timescale, if the temporal fluctuations about that value are small at most times. However, the observable is said to \textit{thermalize} if it equilibrates to the quantum statistical description of the system. In the case of isolated quantum systems, the observables should equilibrate to their longtime averages which should be equivalent to their microcanonical expectation values \cite{gogolin,gge3,Polkovnikov}.

Although there is no fundamental understanding of thermalization in closed quantum systems, it has been proposed that the thermalization for some systems can be understood through {\sl eigenstate-thermalization hypothesis} (ETH)~\cite{deut, sred, alt, Langen}. In other words, non-integrable systems appear to obey ETH \cite{nat}, while numerical studies show that integrable quantum systems equilibrate to a Generalized Gibbs Ensemble (GGE) description \cite{nat,1db,*gge3,*tot,sys2}. More specifically, non-integrable systems at late times relax into a maximal entropic configuration such that energy of the systems remains conserved. This gives rise to the concept of thermalization under ergodic interpretation. On the other hand, an integrable system has many other conserved charges to be worried about, which prevent relaxation into a thermal configuration. With these conserved charges, the maximal entropy configuration turns out to be GGE rather than a true thermal form.

Naturally, there have been rigorous studies in the literature for Fermionic and Bosonic integrable systems. One of the most commonly studied models is the 1-dimensional Fermionic chain that makes a transition from non-integrable to integrable configuration~\cite{Luttinger}. To quantify relaxation in such a system, correlations and distributions are contrasted to the GGE description for various Heisenberg spin configurations~\cite{Guan,*Essler,*Vidmar,*Pozsgay,*Fagotti,*Fagotti2,*Pozsgay2,*Kozlowski,Luca2017,Aschbacher2003,Aschbacher2006,Luca2015}. To investigate GGE, in Ref. \cite{Rigol:0612415}, authors studied hardcore Bosons (HCB) on a lattice where a periodic potential was quenched to drive the system into super-fluid phase. In Refs. \cite{Caneva,*PhysRevA.86.053615}, the authors used a quasi-periodic potential in HCB and showed that the GGE description fails, if translational invariance is broken or there is localization in the system. In Refs. \cite{sys2, Pozsgay2, *Pozsgay3,*Brockmann}, 
the effect of quenching on the ground state (and other interesting states such as N\'{e}el states) of the quasi-periodic potential and description of GGE were investigated. Energetics and transport properties of non-equilibrium states under quench action are also studied extensively \cite{Ogata,Luca2013,Karrasch,Luca2014,Medenjak}.

Reference \cite{1703.09516} recently introduced an alternative approach of truncated truncated GGE (tGGE) in interacting systems. Bosons with repulsive potentials are routinely studied using the Lieb-Linger (LL) model. In Ref. \cite{Caux:PRL2012}, equilibration of integrable or non-integrable systems with quantum quenching in the LL-model have been studied through numerical renormalization group approach. In quantum field theory, the adaptation of the concepts of thermalization has been attempted in Refs. \cite{Mussardo1,*Mussardo2,Doyon:2012bg,Doyon:2014qsa}.

As the reader may have noticed, to overcome the complexities of systematics in the study of thermalization of these systems, most of the studies in the literature have focused on finite-dimensional Hilbert space systems, unless the thermodynamic limit is taken.  All such models, described above, restrict certain features of Bosonic nature for such studies. In this work, we investigate equilibration and thermalization of
systems with \emph{infinite dimensional} Hilbert space, even before going to the thermodynamic limit. For this purpose, we consider Bosonic lattice in a harmonic chain. Interacting bosons on a lattice, favoring certain hopping tendencies have been studied in q-deformed bosons \cite{Pozsgay0, Pozsgay-Eisler} for certain special states. We do not consider them to be in any diverging repulsive potential or ascribe them any hardcore (or for that matter any quantum deformation) properties to rip them off from (or dilute or modify) their fundamental Bosonic property to accumulate. The system we consider consists of two harmonic chains that undergo sudden quantum quench (referred to as quench, hereafter) at an instant of time. The quench action we consider is joining the chains and evolve the chains as one combined chain. Specifically, we investigate the case
where an integrable system makes a jump to another integrable configuration. Thus locally, at a single lattice site, the Hilbert space becomes infinite dimensional and the system is of free bosons (in normal modes) both before and after the quench. In other words, this is a system which remains integrable both before and after the quench. However, the sudden change of a parameter of coupling (from zero to non-zero value) changes the normal modes as well. Since the quench relates one integrable system to another in a linear fashion, the late time configuration can be studied also through Bogoliubov transformation, making it more adaptable to quantum field theoretic settings. We then study the evolution of an initial energy eigenstate for which the
form of late time equilibration is debated. Evolution towards GGE is demonstrated for a system in a non-integrable system quenching to an integrable system \cite{Rigol1}. One of the main motivation for the Harmonic chain is its easy adaptability to field theoretic problems like in non-inertial or uniformly accelerated frames \cite{Mann1,*Mann2,*Mann3}. Therefore, this also allows us to carry the study for all stationary states and in fact to their superpositions as well.

There have been earlier studies for integrable systems like fermionic chains, spin-lattice and the hydrodynamic settings with quenching (see, for instance, Refs. \cite{Antal,*Cstro,*Bertini,*peschel,*eisler,*eisler2008,*caux2013time, *Luca2015,*viti2016,*Rentrop}). Here, we study a Bosonic system under a local quench of joining two sub-chains. We propose a rather new technique to use covariance matrix to check violation of GGE description. We also check density correlation for the system under quench, for verification with the GGE description. It may be noted that density correlations as a check for GGE in LL-model were proposed in Ref. \cite{Nardis}. In Ref. \cite{Langen207}, it was argued that that higher order correlators can be used to demonstrate relaxation to GGE in Bosonic systems. More detailed studies in dynamically evolving extended systems were done through correlations in \cite{Calabrese}.

In our model, we consider the fate of any general state under quench for the infinite dimensional Hilbert space system. We also study the set of states which potentially may defy GGE description. In the literature, three sources of violation of GGE have been identified for a Fermionic system. First is the break down of the invariance of the lattice \cite{sys2,Caneva,Wouters}. Study of interesting aspects of inhomogeneous quenching was done in \cite{Kromos}. Though it has been argued that such violations can be cured by introducing additional conserved charges \cite{Ilievski}. Second, localization plays an important spoilsport \cite{PhysRevA.86.053615} and the third is the divergence of the charges which characterize the GGE description \cite{Kormos}. In such cases, the Lagrange multipliers tend to vanish, effectively removing the memory of conservation. In this work, we try to test these violations while removing the HCB property from the lattice.

In particular, we ask the following questions:
\begin{enumerate}
	\item What happens to the energy eigenstates of {\it an integrable system when the system jumps to another integrable configuration}?
	\item Whether the system settles to an equilibrium state and if yes, to what state the
	system settles to?
	\item whether steady states of pre-quench system develop Gibbs Ensemble characteristics at late times?
	\item What is the measure of the states which exhibit the GGE picture?
\end{enumerate}
We consider both Gaussian (e.g., ground state) and non-Gaussian
(e.g. excited states) initial states for the quenched system.
We study the evolution of such states post-quench,
numerically as well as analytically. We use the properties of Bogoliubov transformation and, in the thermodynamic limit, show that the initial quantum state evolves to GGE form.

If system truly thermalizes, then the system must be completely
described by the first two moments of the distribution in phase space~\cite{deut, sred}. Hence, in this work, we focus on the structure of the covariance matrix to identify a necessary (not sufficient) condition for the system to be thermalized. Using this, we classify the set of initial states which will (or will not) evolve to a thermal configuration. Our analysis also provides a possible link of thermalization with delocalization in the Hilbert space similar in spirit to the classical equivalent --- delocalization in the phase space for classical thermalization. Using these schemes we verify that for Bosonic systems, GGE description remains {\it largely} valid for states for which delocalization is large (studied through inverse participation ratio). However, this scheme also allows us to find states which are delocalized yet the GGE description fails, prominently because of divergence of the conserved charge. Thus our study delinks the localization in a Bosonic system from the divergence of the charges.

\section{\label{sec:model}THE MODEL and setup}

The system we consider consists of two harmonic chains, with lattice sizes $N$ and $M$, respectively and use static boundary conditions for each block, which means $q_{1}=q_{N \text{ or } M}=0$, with $q_i$ depicting the oscillator at $i-$th lattice site. For large $N$, the results are independent of the choice of the boundary condition. For $t<0$, the two chains are non-interacting whose Hamiltonian is $H_0 = H_N+H_M,$ with,
\begin{equation}
	H_{K} = \sum_{i=1}^{K} \frac{p_i^2}{2 m} + \frac{1}{2}m \omega_0^2(q_{i+1}-q_{i})^2 \quad K = N~\mbox{and}~M \, \, \end{equation}

At $t=0$ we turn on an interaction between the two chains with coupling strength $\omega_0$.  Post quench $(t\geq0)$, the  Hamiltonian is $H = H_0 + H_{int}$ where $H_{int}= -m \omega_0^2 q_{N+1} q_N$.  In terms of creation and annihilation operators, the Hamiltonian operators are,
{\small	
	\begin{eqnarray}		
		\hat{H}_0 &=& \sum_{l=1}^{N} \hbar\ \omega_l \left(\hat{a}^\dagger_l \hat{a}_l
		+ \frac{1}{2}\right) + \sum_{l=1}^{M} \hbar \omega_{_{N+l}}
		\left(\hat{a}^\dagger_{_{N+l}} \hat{a}_{_{N+l}} + \frac{1}{2}\right) \nn \\
		\label{quenchmodefin}
		\hat{H}&=&\hat{H}_N+\hat{H}_M+\hat{H}_{int} = \sum_{l=1}^{N+M} \hbar \omega_l (\hat{c}^\dagger_l  \hat{c}_l + \frac{1}{2}) \, ,		
	\end{eqnarray}	
}
where $\omega_l$'s are normal mode frequencies (see Appendix \ref{App:A}).

Let the system be in an eigenstate $|\psi\rangle_i$ of the Hamiltonian $\hat{H}_0$. After quench, the time evolution of the system is described by the Hamiltonian \ref{quenchmodefin}. 
In order to compute the time evolution, we need to map $|\psi\rangle_i$ to a linear combination of the eigenstates of $\hat{H}$. Hence, to express $|\psi\rangle_i$ in terms of the eigenstates of $\hat{H}$  we need to,

\begin{enumerate}
	\item Get a relationship between the creation and annihilation operators before and after the quench, and	
	\item Express the ground state $|0\rangle_1 = |0\rangle_N \otimes |0\rangle_M$ of the disjoint system as a linear combination of an eigenstate of the joint system.
\end{enumerate}
Other eigenstates can be obtained through the action of $\hat{a}^\dagger_l$ on $|0\rangle_1$.  

We can relate creation and annihilation operators before and after the quench using the Bogoliubov transformations (see Appendix \ref{App:B}). It is in general non-trivial to relate the ground state ($|0\rangle_N \otimes |0\rangle_M$) corresponding to $\hat{H}_0$ to the ground state corresponding to $\hat{H}$. 

However, using the fact that the ground state is a Gaussian for the integrable systems, one can write the following general expression connecting the two: 
\begin{equation}
	\label{ansatz}
	|0\rangle_N \otimes |0\rangle_M = N_0 e^{-\frac{F_{ij}}{2}\hat{c}^\dagger_i\hat{c}^\dagger_j + G_{ij} c_i c^{\dagger}_j + \cdots} |0\rangle_{N+M}
\end{equation}
where $F_{ij}, G_{ij}$ are  $(N+M) \times (N+M)$ matrices. It can be seen that only $F_{ij}$ contributes to the expectation values of the physical quantities at the leading order of expansion. In order to do numerical computations, we need to take finitely many terms from the state expansion $\exp(\hat{O}) = 1 + \hat{O} + \hat{O}^2/2 + \cdots$, i.e. we need to collect contributions of different occupancy states in the new basis. We can see through inverse participation ratio in \ref{tabdeloc} and \ref{graph:IPR}, contributions from higher states steadily fall down with larger lattice size. 

$F_{ij}$ can be determined by solving the following constraint equations,

\begin{eqnarray}	
	\label{constr}
	\hat{a}_i |0\rangle_N \otimes |0\rangle_M = 0, & & \forall \, i =	
	1,2,\ldots (N + M ).	
\end{eqnarray}

We expect $F_{ij}$ to be symmetric since $\hat{c}_l^\dagger$ 
operators commute with each other.  Using these results  and \ref{ansatz},  any generic eigenstate of $\hat{H}_0$ is given by
\begin{equation}
	\label{restate}
	|\psi\rangle_{\{(n_i,n_j)\}} = \prod_{i=1}^{N + M }
	\prod_{j=1}^{M}(\hat{a}^\dagger_i)^{n_i} e^{-\frac{F_{lk}}{2}\hat{c}^\dagger_l\hat{c}^\dagger_k} |0\rangle_{N+M}.
\end{equation}

Using \ref{restate} and the Bogoliubov transformations in Appendix \ref{App:B}, we can therefore write this state as a linear combination of eigenstates of the Hamiltonian $\hat{H}$.

Numerically, we can study the time evolution with a finite number of the basis states, hence, we restrict \ref{restate} to the first order expansion of the exponential term. To study equilibrium relaxation of the system, we need to look at physical observables associated with it. We use the occupation number of the modes in one part of the disjoint system,
\begin{eqnarray}	
	\label{oper1}	
	\hat{n}_m &=& \hat{a}^\dagger_m \hat{a}_m,\ m=1,2\ldots,(N)
\end{eqnarray}

These are the conserved charges of the system and also determine the energy of the system prior to the quench.  It is important to note that our interest is to consider large lattice limit in the large time limit. So mostly, we are interested in large time behaviour of the system post-quench. However, from the point of view of the further application of our results in infinite dimensional Hilbert space, we will also be studying the system in the usual thermodynamic limit (where the number of lattice sites being extremely large). Since these operators do not commute with the Hamiltonian after the quench, the time evolution is non-trivial. We also consider correlation of the number operators. Further our scheme of thermodynamical limit to be taken post the late time limit is different from what is typically done \cite{Aschbacher2003,Aschbacher2006,Bernard,Bernard2015,Bernard2016}. Since we wish to explore a system which settles down to an equilibrium configuration (at least) in the late time (by making it 
asymptotically free or integrable) and analyze what is the end equilibrium configuration as the number of degree of freedom grows, we will first let the system evolve into a configuration under the action of quenching and will then take the large $N$ limit. This will also enable us to visualize the field theoretic generalization in time evolving scenarios more accurately, when a system asymptotically relaxes in an integrable configuration. This quantum field theoretic exercise we will pursue elsewhere.

One other aspect of this strategy is to reflect upon the nature of the Lagrange multiplier at the late time limit, which will be more justified when the post-quench system remains non-integrable. Thereafter, the evolution of the energies of the disjoint modes is, 
{\small
	\begin{eqnarray}
		\label{eq:Observables}
		E_N(t) &=& \sum_{i=1}^{N} \left[\hat{n}_i(t) + \frac{1}{2} \right] \hbar \omega_i; \\
		E_M(t)&=& \sum_{i=1}^{M} \left[\hat{n}_{N+i}(t) + \frac{1}{2} \right] \hbar \omega_{N+i}. \nonumber
	\end{eqnarray}
}

\section{\label{sec:results}Numerical RESULTS}

\begin{figure}[!htb]
	\centering
	\includegraphics[height=0.4\linewidth,width=\linewidth]{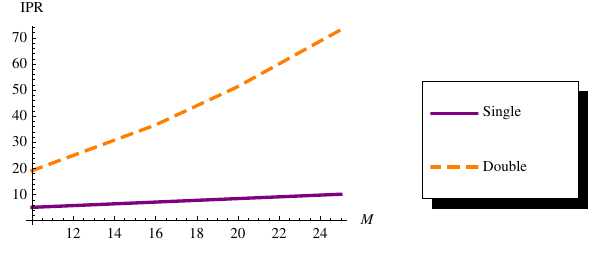} \caption{IPR of different states and lattice size}
	\label{graph:IPR}
\end{figure}

We begin the system in an eigenstate of the disjoint Hamiltonian ($\hat{H}_0$). Once the system is quenched, the eigenstate can be written as linear combination of the normal mode eigenstates of the quenched Hamiltonian  --- \emph{delocalization} in the Hilbert space of the quenched Hamiltonian. \ref{tabdeloc} gives the delocalization for different initial states and lattice sizes (for $N = 5$ and different $M$), where $\mathcal{N}[\psi_i]$ is the approximate number of eigenstates involved in writing the first order approximation of \ref{restate}. Delocalization in phase space has previously been studied from the point of view of thermalization \cite{sys1} and  inverse participation ratio ($IPR = 1/\sum_i p_i^2,$) \cite{Murphy}, where $p_i$ are the probability amplitudes of finding the state in post-quench eigenstate $|i\rangle$. \ref{graph:IPR} contains the plot of IPR for different states and lattice sizes.

\begin{table}[!h]	
	\begin{center}	
		\begin{tabular}{|c|c|c|c|| c| c| c |c|c|}
			\hline  M &  $|\psi_i\rangle$ & $\mathcal{N}[\psi_i]$ & IPR &  M &  $|\psi_i\rangle$ & $\mathcal{N}[\psi_i]$ & IPR\\ 
			\hline  10 & $a^\dagger_i$ & 695 & 5.07118&  10 &  $a^\dagger_3a^\dagger_4$ & 3181 &19.1345 \\ 			
			\hline 16 &  $a^\dagger_i$ & 1791 &  7.09398 &  16 &  $a^\dagger_3a^\dagger_4$ & 10858 &  36.6371 \\ 
			\hline  20 &  $a^\dagger_i$ & 2950 & 8.43266 &  20 &  $a^\dagger_3a^\dagger_4$ & 20801 &  51.3287  \\ 	\hline 
		\end{tabular} 
	\end{center}
	\caption{Table shows the delocalization of initial state $\psi_i$ for $N = 5$ and different $M$ lattice sites.}
	\label{tabdeloc}
\end{table} 

\vspace*{5pt}

\begin{figure}[!htb]
	\centering
	\includegraphics[width=0.95\linewidth]{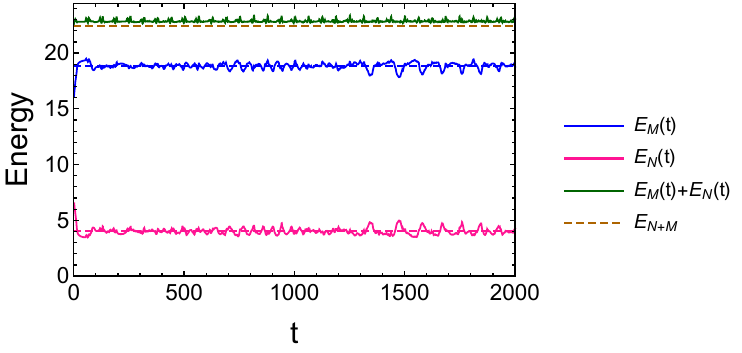}
	\caption{For the initial state $|\psi_i\rangle = a^\dagger_3 a^\dagger_4 |0\rangle_N \otimes |0\rangle_M $ with $N=5$ and $M = 25$. $E_N(t)$ (in pink) and its long time average , $E_M(t)$ (in blue) and its long time average, the total energy of the state after the quench (in green) and $E_N(t)+E_M(t)$ (in yellowish green) }
	\label{graph:evol2} 
\end{figure}

\noindent {\it Equilibration, steady states and fluctuations:} In order to go about understanding equilibration, we evaluate the observables $E_N(t), E_{M}(t), E_N(t) + E_{M}(t)$ as a function of time. From \ref{graph:evol1} and \ref{graph:evol2}, we infer the following: 
\begin{figure}[!htb] 
	\centering
	\includegraphics[width=0.95\linewidth]{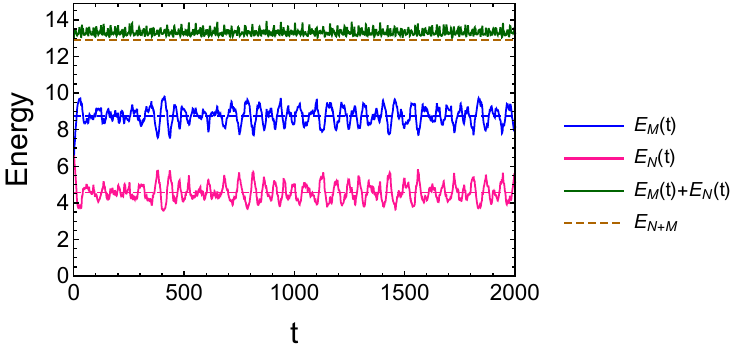}
	\caption{For the initial state $|\psi_i\rangle = a^\dagger_3 a^\dagger_4 |0\rangle_N \otimes |0\rangle_M$ with $N=5$ and $M=10$. $E_N(t)$ (in pink) and its long-time average, $E_M(t)$ (in blue) and its long time average, the total energy of the state after the quench (in green) and $E_N(t)+E_M(t)$ (in yellowish green)}	
	\label{graph:evol1} 
\end{figure}

\begin{figure}[!htb]
	\centering
	\includegraphics[width=0.95\linewidth]{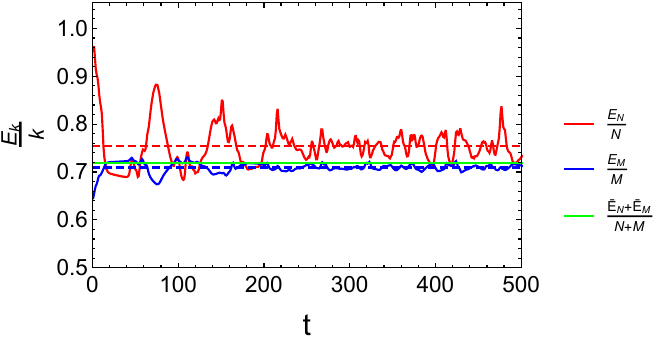}\\
	\caption{$E_N/N$ and $E_M/M$ along with their long time averages for t=(0,500) time steps for lattice sizes $N=5$, $M=20$ in initial state $|\psi_i\rangle = a^\dagger_3|0\rangle_N \otimes |0\rangle_M $}
	\label{graph:sub520-1}
\end{figure}

\begin{figure}[!htb]
	\centering
	\includegraphics[width=0.95\linewidth]{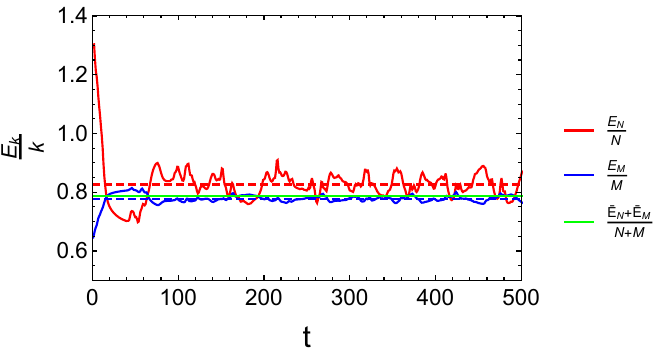}
	\caption{$E_N/N$ and $E_M/M$ along with their long time averages for t=(0,500) time steps for lattice sizes $N=5$, $M=20$ in initial state $|\psi_i\rangle = a^\dagger_3|0\rangle_N \otimes |0\rangle_M $}
	\label{graph:sub520-2}
\end{figure}

As we increase the lattice size and the number of excitation modes in the initial state, eigenstates are more \emph{delocalized} (i.e. smaller IPR)(see \ref{graph:IPR}) in the Hilbert space of the quenched Hamiltonian.  As we increase the lattice size --- corresponding to increase in delocalization of the initial states --- the expectation value of the energy tends towards the long-time average.  This is similar to the notion of equilibration in classical systems where two systems at different temperatures in contact, at late times, reach the same temperature. In this case, the two chains are in two different energy eigenstates, however, at late times, both the systems tend to a state with similar average energy per mode (as can be seen in \ref{graph:sub520-1} and \ref{graph:sub520-2}) in the relaxation  timescales of the system.  In this sense, the equilibration (late time relaxation) is equivalent to the equipartition as well. Equilibration is associated with a late time relaxation to a 
steady configuration.

\begin{figure}[!htb]	
\centering	
	\includegraphics[width=0.8\linewidth]{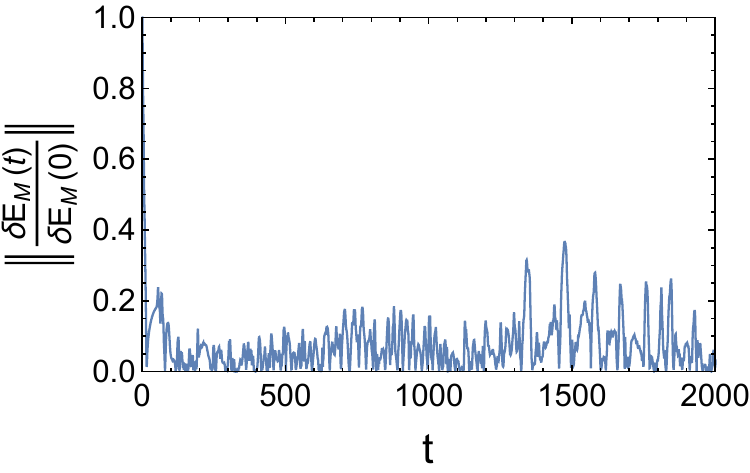}
	\caption{Plot of the fluctuations for initial state $|\psi_i\rangle = a^\dagger_3 a^\dagger_4 |0\rangle_N \otimes |0\rangle_M$ with $N = 5$ and $M = 25$ for t= [0,2000] for different lattice sizes.} \label{graph:fluc1-2}
\end{figure}

\begin{figure}[!htb]	
\centering
\includegraphics[width=0.8\linewidth]{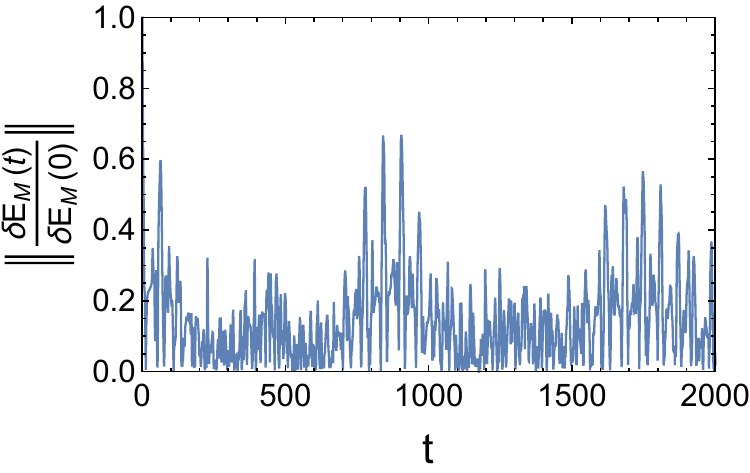}
\caption{For lattice sizes $N=5$, $M=16$ with initial state $ |\psi_i\rangle = a^\dagger_3 |0\rangle_N \otimes |0\rangle_M$. Plot of $\frac{||\delta E_M(t)||}{||\delta E_M(0)||}$ for t=(0,2000) time steps}
\label{graph:fluc2-1}
\end{figure}

\begin{figure}[!htb]
	\includegraphics[width=0.8\linewidth]{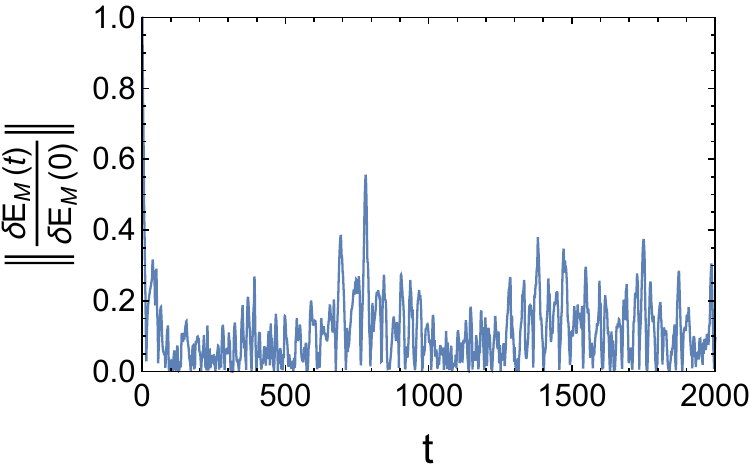}
		\caption{For lattice sizes $N=5$, $M=16$, with initial state $|\psi_i\rangle = a^\dagger_3 a^\dagger_4 |0\rangle_N \otimes |0\rangle_M$. Plot of  $\frac{||\delta E_M(t)||}{||\delta E_M(0)||}$ for t=(0,2000) time steps}
	\label{graph:fluc2-2}
\end{figure}

To further investigate, let us define, 
\bea
\frac{||\delta E_M(t)||}{||\delta E_M(0)||} =\frac{||E_M(t)-\overline{E_M}||}{||E_M(0)-\overline{E_M}||}, 
\eea 

which is the time evolution of the fluctuations of $E_M$ about the  long time average value of observable $H_M$.  From \ref{graph:fluc1-1}, \ref{graph:fluc1-2}, \ref{graph:fluc2-1} and \ref{graph:fluc2-2}, we infer the following:
\begin{figure}[!htb]
\centering
	\includegraphics[width=0.8\linewidth]{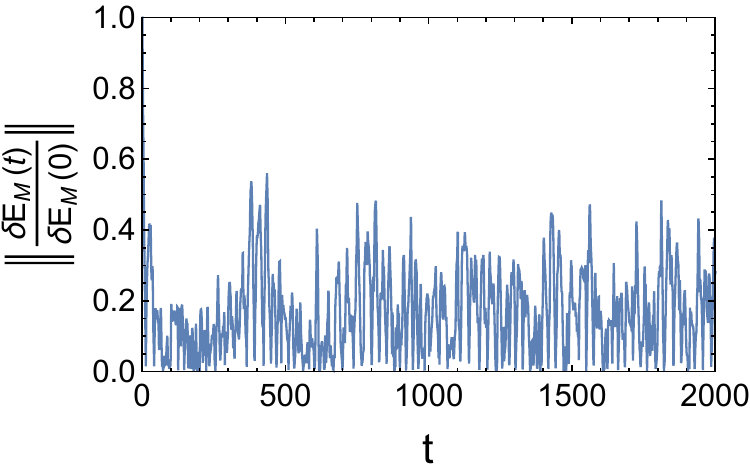}
	\caption{Plot of the fluctuations for initial state $|\psi_i\rangle = a^\dagger_3 a^\dagger_4 |0\rangle_N \otimes |0\rangle_M$ with $N = 5$ and $M = 10$ for t= [0,2000] for different lattice sizes.}
	\label{graph:fluc1-1}
\end{figure}


The fluctuations remain subdued periodically. As the lattice size increases, this recurrence is rarer. Also for the same lattice, there are smaller fluctuations with farther recurrences for initial states which have better delocalization (See \ref{tabdeloc}). The quasi-periodic nature of observables in finite systems imply that there may be recurrence in long times \cite{venuti,tot}, however, the time it takes to relax to a steady state from a non-equilibrium initial condition should be much less than the recurrence timescale of the system. As seen here, the recurrence timescale grow with the system size and it is expected that it grows exponentially with the system size~\cite{yukalov,extra}. Hence, recurrences become rarer in the thermodynamic limit.  

To understand this further and to know whether we can define \emph{equilibration} between two quantum subsystems, in \ref{graph:sub520-1} and \ref{graph:sub520-2}, we plot the evolution of the average energy per mode for the two subsystems, $E_N/N$ and $E_M/M$. We observe the following salient features: 
\begin{itemize}	
	\item The longtime averages of the observables (per lattice point) relax to GGE expectation value.
	\item The fluctuations about the average value become smaller with increasing lattice size of the system.
	\item The per-point observable quantities approach a common value for a specific realization (system size).
\end{itemize}

\section{\label{sec:analytical}Analytical Understanding}

The numerical evolution of the states clearly indicates that the system relaxes to GGE. However, the numerical analysis does not provide a physical understanding as to (i) why the long time average of the observables relax to GGE?  (ii) For what initial conditions, thermalization is violated? and (iii) Is there a way to identify the relation between the delocalization in the Hilbert space and equilibration of states? To go about understanding these issues, we resort to analytical methods and study the observable characteristics of the system at late times. 

For the system of Harmonic chains, the normal mode analysis (see Appendix \ref{App:A}) makes the study of the system as that of independent oscillators. For this system, prior to the quench, the ground state is a multi-mode Gaussian state. As the quench action is also ``quadratic'', the time evolution keeps the state to be  Gaussian although the normal modes of the system before and after quench are different. Therefore, the expectation values of any operator can be obtained from the Gaussian density matrix prior to or post quench, with appropriate Bogoliubov coefficients (See Appendix \ref{App:B}).

In Appendices \ref{App:C} and \ref{App:D}, using Bogoliubov transformations, we obtain the evolution and the long time averages of these observables. Using the results in  Appendices (\ref{App:B}, \ref{App:C} and \ref{App:D}), we can rewrite the conserved quantities (through the average expectation of the post quench number operator) in terms of the Bogoliubov coefficients as

\begin{equation}
	\langle\hat{n'}_k \rangle = \sum_l |\beta_{k l}|^2 +\sum_j\left(|\alpha_{k j}|^2 n_j+ |\beta_{k j}|^2 n_j\right).  \label{Emission} 
\end{equation}

Physically, the first term in the RHS corresponds to the occupancy when the initial state is vacuum state. This is referred to as \emph{vacuum polarization}~\cite{Birrell} under the quench. The remaining terms are from the occupancy of the initial states prior to quench, and is referred to as {\it stimulated emission} \cite{Muller}.

Having obtained the results for a general initial state,  we can go about understanding the above questions. 

First, the periodicity observed in the expectation values directly stem from the beat frequencies of the normal modes. As can be seen from Appendix \ref{App:A}, in the thermodynamic limit, the beat frequencies become smaller and the recurrences become rarer. Second, setting the initial state to be ground state, we compute the long-time average of the observables and compare them with the thermal states.

As mentioned earlier, GGE can be used to describe the steady states for integrable systems evolving from a non-equilibrium initial state. With such a description one can generalize the concept of thermalization for integrable systems to relaxation into GGE. We would like to, therefore, compute GGE expectation values of the observables mentioned above and thereby verify the validity of such a description for these states.

GGE is completely described in terms of conserved quantities of the system. Therefore, for our system, we need to find them post quench. The energies (and hence occupation number) of each normal mode in the  joint system (after the quench) are conserved as there will be no interactions between these modes and the system is similar to a system of $N + M$ independent harmonic oscillators, with the independent conserved quantities $\hat{n'}_k =  \hat{c}^\dagger_k \hat{c}_k,\ k=1,2,\ldots,(N+M)$. Thus, the density matrix of  GGE is,

\begin{equation}	
	\label{GGEsys}
	\hat{\rho}_{GGE} = \frac{e^{-\sum_{m=1}^{N+M}	
			\lambda_m \hat{n'}_m}}{Tr(e^{-\sum_{m=1}^{N+M}
			\lambda_m \hat{n'}_m})}
\end{equation}

where $\{\lambda_m\}$ are Lagrange multipliers.

We can find the Lagrange multipliers, using the initial 
conditions of the conserved quantities,

\begin{equation}		
	\label{ggeconsv}
	Tr(\hat{\rho}_{GGE} \hat{n'}_k)=\langle\hat{n'}_k(0)\rangle
\end{equation}

to be 

\begin{equation}	
	\label{lagrange}	
	\lambda_k =	
	\ln\left(\frac{1+\langle\hat{n'}_k(0)\rangle}{\langle\hat{n'}_k(0)\rangle}\right)	
\end{equation}

for $k=1,2\ldots (N+M)$. Therefore, the Lagrange's multipliers are obtained from the Bogoliubov coefficients. Also, once the state $|\psi_i\rangle$ is known in terms of \ref{restate}, the conserved quantities are known  and the Lagrange multipliers are fixed based on the initial state. Vice versa, one can try to obtain the information regarding the initial state through the GGE parameter at late time equilibrium.

This feature is reminiscent of the integrability of the system both pre and post-quench. The late time configuration has to respect the conservation of charges, however, as we will see below if the transformation between the pre quench integrable configuration and post-quench integrable configuration are linearly related aka related through Bogoliubov transformations, the late time conserved charges are totally expressible in terms of early time conserved quantities the state dependency survives for non vacuum initial states \cite{Lochan2016}. However, as we can see that with divergent conserved charges $\langle\hat{n'}_k(0)\rangle$, the appearance of $\lambda$ in the  GGE gets weaker.

In a quantum field theoretic generalization \cite{Mussardo1,*Mussardo2,Doyon:2012bg,Doyon:2014qsa} of the idea of quenching, it will be interesting to see the differentiability of the GGE equilibration with thermalization. Further, in interacting field theories, such differentiability can be dynamically studied \cite{1703.09516}.

Operators $\hat{c}$ and $\hat{c}^\dagger$ are traceless matrices in the energy eigenbasis and the trace of a tensor product will be a product of the traces of the individual components. This implies that the correlation matrices   $\langle\hat{c}_i\hat{c}^\dagger_j\rangle_{GGE} =\delta_{i,j} \langle\hat{c}_jc^\dagger_j(0)\rangle$ and $\langle c^\dagger_ic_j\rangle_{GGE} =\delta_{i,j} \langle\hat{c}^\dagger_j\hat{c}_j(0)\rangle$ will be diagonal and $\langle\hat{c}_i\hat{c}_j\rangle_{GGE}=\langle\hat{c}^\dagger_i\hat{c}^\dagger_j\rangle_{GGE}=0$. Hence, using this and \ref{ggeconsv},

\begin{widetext}	
	\begin{equation}		
		\label{opergibbs1}		
		\begin{aligned}			
			\langle\hat{n}_m(t)\rangle_{GGE}			
			=&\frac{4}{(N+M+1)(N+1)}			
			\sum_{i=1}^{N}\sum_{j=1}^{N}\sum_{k=1}^{N+M}			
			\bigg[\cosh^2(\gamma_{_{mk}})\langle\hat{c}^\dagger_k	
			\hat{c}_k(0)\rangle +\sinh^2(\gamma_{_{mk}})			
			\langle\hat{c}_k			
			\hat{c}^\dagger_k(0)\rangle\bigg]\\ & \times			
			\sin(\frac{\pi j k}{N+M+1})\sin(\frac{\pi j				
				m}{N+1}) \sin(\frac{\pi i k}{N+M+1})			
			\sin(\frac{\pi i m}{N+1})			
		\end{aligned}		
	\end{equation}
	
	\begin{equation}	
		\label{opergibbs2}	
		\begin{aligned}		
			\langle\hat{n}_{N+m}(t)\rangle_{GGE} =&		
			\frac{4}{(N+M+1)(M+1)}		
			\sum_{i=1}^{M}\sum_{j=1}^{M}\sum_{k=1}^{N+M}		
			\bigg[\cosh^2(\gamma_{_{N+m,k}})\langle\hat{c}^\dagger_k		
			c_k(0)\rangle\\ &+\sinh^2(\gamma_{_{N+m,k}})\langle\hat{c}_k		
			\hat{c}^\dagger_k(0)\rangle\bigg]\times		
			\sin(\frac{\pi (N+j)			
				k}{N+M+1})\\ &\times\sin(\frac{\pi j			
				m}{M+1}) \sin(\frac{\pi (N+i) k}{N+M+1})		
			\sin(\frac{\pi i m}{M+1})		
		\end{aligned}	
	\end{equation}
\end{widetext}

Since, $\hat{c}^\dagger_i\hat{c}_j$ (and so $\hat{c}_i\hat{c}^\dagger_j$ ) are conserved quantities and do not evolve in time we can immediately conclude that \textit{long time averages} computed in Appendix \ref{App:D} match the expectation values from the Gibbs Generalized Ensemble.

For thermal states, the evolution in phase space can be described (by virtue of Gaussianity of the density matrix or Wigner function) by the mean and the second moments. Furthermore, the form of the thermal covariance matrix is tightly constrained. Not only it should be diagonal, but the diagonal entries tell us about the Gibbs conserved charges \cite{Olivares}, that can be used to compare against the numerical relaxation values. The analysis shows that the Gaussian state indeed relaxes to  GGE. In Appendix (\ref{App:F}),  we verify the same using the covariance matrix analysis. 

Third, covariance matrix analysis is not sufficient in dealing with non-vacuum energy eigenstates, as these states are not Gaussian. In the case of non-vacuum initial states, as seen from \ref{Emission}, the observables will receive correction over the GGE description. These corrections can be characterized as arising from stimulated emissions. In Appendix (\ref{APP:G}), we do a detailed analysis on such stimulated emission induced corrections. We observe that for larger lattice size, these corrections are suppressed as $1/N$, where $N$ is the lattice size.  In order to verify the description of the system post quench at higher orders of operator expectation, we verify the lattice correlation against its anticipated GGE description in Appendix (\ref{APP:H}). Further, lower the initial excitation is, smaller is the deviation from GGE. In the thermodynamic limit, the per-point observable quantities get effectively described by GGE (vacuum description), albeit with different values of the parameters in the 
vacuum state (due to the $N$ dependence). Again, we can identify the generic states which will show significant departures from GGE description. We show that there exists set of well-delocalized states which have the tendency to disobey the GGE description. This behaviour can be understood through divergent conserved charges (in the limit of $\langle \hat{n}'_k(0) \rangle \to 0)$ which make (some of) the Lagrange multipliers \ref{lagrange} vanish thus letting the description handicapped with the insufficient number of parameters to capture the conservation.

\section{\label{sec:disco}CONCLUSIONS AND DISCUSSION}

While the mechanism and validity of \emph{generalized thermalization} in integrable systems --- where ETH is not valid is still not well understood ---  many numerical studies have been done with spin systems to verify the validity of the GGE description of their steady states \cite{nat, sys2, 1db, extra}.  There have been studies investigating whether a sudden change in parameters of the Hamiltonian of an integrable system to another leads to GGE. These studies, however, are mostly done for finite dimensional Hilbert space systems. The harmonic oscillator chain with a finite lattice size, that we have considered, is an infinite dimensional Hilbert space system. So a non-equilibrium initial state can delocalize (i.e. larger IPR values) into infinitely many possible combinations of independent states when we take it into an out of equilibrium condition. Our system also quenches between two integrable configurations in a linear fashion. Therefore, we can utilize the techniques of Bogoliubov transformations for 
our study. We have investigated the \emph{relaxation} of eigenstates of such a system.

We have verified the validity of the GGE to describe the long-time average values starting from a non-equilibrium initial state. We also explicitly showed the fluctuations about the steady state with the time evolution of this system. For this system, the longtime averages match the GGE expectation values, yet the fluctuations about this long time average value, although small in the relaxation timescales, are still not negligible. 

We were also able to see a form of \emph{equilibration between the subsystems} for our system. It appears as though the two subsystems are approaching a limit in which the average energy per excitation modes are the same within the relaxation time scales, similar to classical systems. If for a given initial eigenstate, the system indeed evolves to GGE description, the characteristics describing system should be identified and compared against those of GGE. Under general arguments of stimulated emission between integrable systems and thermodynamic limit, we obtained GGE description for finite excitations as well. Using covariance matrix and methods of stimulated emission, we showed that the evolution of the general state (not only eigenstates) to a GGE form in the thermodynamic limit. By corollary, we have obtained the states which may potentially not evolve into a GGE form. 

The measure of such states in the Hilbert space is an interesting quantity in order to comment on the generic feature of equilibrium relaxation. The results obtained here can also be attempted for scenarios where the system evolves from one integrable configuration to another post quench, gradually in an asymptotic way, which makes the study interesting from many quantum field theoretic scenarios. We will pursue these issues in a subsequent work.

\section{Acknowledgement}
Computaions are done using {\sl Quantum Package} in Mathematica.
The authors wish to thank Arul Lakshminarayan, Krishnanand Mallaya and Sandra B for useful
discussions. The work is supported under DST-Max Planck Partner Group
on Cosmology and Gravity. SM was supported by Inspire fellowship of
DST, Government of India. Research of KL is supported by DST-INSPIRE Faculty Fellowship of the Government of India.
The authors thank the referees for their constructive suggestions which helped improving the manuscript substantially.
\appendix
       
	\section{The Harmonic chain}
	\label{App:A}
	A harmonic oscillator chain is a well known integrable
        system. With periodic boundary conditions, this Hamiltonian in
        the thermodynamic limit would correspond to a free field
        theory. The Hamiltonian of a N body harmonic oscillator
        lattice with nearest neighbor interactions can be written as,
	
	\begin{equation}
		H=\sum_{i=1}^{N} \frac{p_i^2}{2 m} + \frac{1}{2}m
                \omega_0^2(q_{i+1}-q_{i})^2
	\end{equation}
	\noindent where $\omega_0$ is the natural frequency of the
        individual oscillator, $p_i$ and $q_i$ are the respective
        momentum and position coordinates for a harmonic oscillator on
        the $i^{th}$ lattice site with a lattice distance of $a$. In
        this work we use static boundary conditions, which
        means $q_{1}=q_{N+1}=0$. For large $N$, the results are
        independent of the choice of the boundary condition.
	
	Rewriting the Hamiltonian in terms of the normal modes leads to
	a system of N uncoupled harmonic oscillators:
	\begin{equation}
		\hat{H} = \sum_{k=1}^{N} \hbar\ \omega_k
                (\hat{a}^\dagger_k \hat{a}_k + \frac{1}{2})
	\end{equation}
	where $\hat{a}^\dagger_k$ and $\hat{a}_k$ are the creation and
        annihilation operators with normal mode frequencies,
	\begin{equation}
		\omega_k = 2 \omega_0 \left| \sin\left(\frac{\pi k
                }{2(N+1)}\right) \right|.
	\end{equation}
	As we will explicitly calculate in other sections of the appendix, the beat frequencies appearing in the various expectations are determined by 
	\bea
	\left|\omega_k - \omega_j\right| &=& 2 \omega_0 \left| \sin\left(\frac{\pi k}{2(N+1)}\right) - \sin\left(\frac{\pi j}{2(N+1)}\right)\right| \nonumber \\
	 &=& 4 \omega_0 \left|\sin\left(\frac{\pi (k-j)}{2(N+1)}\right)\cos\left(\frac{\pi (k+j)}{2(N+1)}\right)\right|.
	\eea
	The  $(k-j)$ takes values $N+1 -m$ with $m \in 2,3,..$. Therefore, in the thermodynamic limit, most of the
	beat frequencies become smaller and therefore the recurrences occur lately. For smaller lattice sizes, two normal mode 
	frequencies can easily be separate enough to show periodicity effectively, while for the larger systems,
	the probability of finding separate enough normal modes will be rarer, hence the oscillatory terms get suppressed in overall contribution.
	
	\section{Bogoliubov Transformations}
	\label{App:B}
	
	As mentioned earlier, the normal modes of the system before
        and after the quench are different. In order to obtain the
        relation between the creation and annihilation operators
        before and after the quench, we need to diagonalize the
        Hamiltonian by applying the Bogoliubov transformations.  Using
        the definitions for normal mode coordinates before and after
        the quench ($Q_k$ and $Q'_k$ respectively), we have,
        {\small
	\begin{eqnarray}
		Q_l &=& \sqrt{\frac{2}{N+1}} \sum_{m=1}^{N} q_m
                \sin\left(\frac{\pi l m}{N+1}\right) = \sqrt{\frac{4}{(N+1) (N+M+1)}} \nonumber\\ &\times&
                \sum_{m=1}^{N} \sum_{k=1}^{N+M}Q'_k \sin
                \left(\frac{\pi m k}{N+M+1} \right)
                \sin\left(\frac{\pi l m }{N+1}\right), \label{Q2PQ} 
	\end{eqnarray}
	}
	where $q_k$ is real space variable.
	We can use the relationship of the creation and annihilation
        operators before and after the quench to get,
	
	\begin{eqnarray}
		\label{bogostep}
		\hat{a}_l + \hat{a}^\dagger_l &=&
                \sqrt{\frac{2}{N+1}}\sqrt{\frac{2}{N+M+1}}\sum_{m=1}^{N}
                \sum_{k=1}^{N+M} \\ &\times&
                e^{\gamma_{_{l,k}}}(\hat{c}_k +
                \hat{c}^\dagger_k) \sin \left(\frac{\pi m k}{N+M+1}
                \right) \sin\left(\frac{\pi l m }{N+1}\right) \nonumber
	\end{eqnarray}
	where $e^{\gamma_{_{l,k}}} =
        \sqrt{\frac{\omega_l}{\omega'_k}}$. 
        
        Similarly we can use $P'_l$ to get an expression for $\hat{a}_l - \hat{a}^\dagger_l$ and adding
        these to expressions lead to:
	\begin{widetext}
	\begin{equation}
		\begin{aligned}
			\label{bogo1}
			\hat{a}_l =
                        \sqrt{\frac{2}{N+1}}\sqrt{\frac{2}{N+M+1}}\sum_{m=1}^{N}
                        \sum_{k=1}^{N+M} \left( \cosh(\gamma_{_{l,k}})
                        \hat{c}_k + \sinh(\gamma_{_{l,k}})
                        \hat{c}^\dagger_k\right) \sin
                        \left(\frac{\pi m k}{N+M+1} \right)
                        \sin\left(\frac{\pi l m }{N+1}\right)
		\end{aligned}
	\end{equation}
	The Hermitian conjugate of \ref{bogo1} lead to the
        expansion for $\hat{a}^\dagger_l$.  Similarly for $\hat{a}_{N+l}$
        we get,
	\begin{equation}
		\begin{aligned}
			\label{bogo2}
			\hat{a}_{N+l} =
                        \sqrt{\frac{2}{M+1}}\sqrt{\frac{2}{N+M+1}}\sum_{i=1}^{M}
                        \sum_{k=1}^{N+M} \left(
                        \cosh(\gamma_{_{N+l,k}}) \hat{c}_k +
                        \sinh(\gamma_{_{N+l,k}})
                        \hat{c}^\dagger_k\right) \sin
                        \left(\frac{\pi (N+i) k}{N+M+1} \right)
                        \sin\left(\frac{\pi l i }{M+1}\right)
		\end{aligned}
	\end{equation}
	Thus, we obtain the \textit{Bogoliubov}
        transformations that relate the creation and annihilation
        operators of the system before and after the quench as
        \bea
        \alpha_{lk} &=& \sqrt{\frac{2}{N+1}}\sqrt{\frac{2}{N+M+1}}\sum_{m=1}^{N}
                         \cosh(\gamma_{_{l,k}})\sin
                        \left(\frac{\pi m k}{N+M+1} \right)
                        \sin\left(\frac{\pi l m }{N+1}\right),\\
         \beta_{lk} &=& \sqrt{\frac{2}{N+1}}\sqrt{\frac{2}{N+M+1}}\sum_{m=1}^{N}
                         \sinh(\gamma_{_{l,k}})\sin
                        \left(\frac{\pi m k}{N+M+1} \right)
                        \sin\left(\frac{\pi l m }{N+1}\right)  \label{Bogol1}
         \eea
	for $l:1\rightarrow N$, and
	\bea
        \alpha_{lk} &=&  \sqrt{\frac{2}{M+1}}\sqrt{\frac{2}{N+M+1}}\sum_{i=1}^{M}
                         \cosh(\gamma_{_{N+l,k}}) \sin
                        \left(\frac{\pi (N+i) k}{N+M+1} \right)
                        \sin\left(\frac{\pi l i }{M+1}\right),\\
         \beta_{lk} &=& \sqrt{\frac{2}{M+1}}\sqrt{\frac{2}{N+M+1}}\sum_{i=1}^{M}
                         \sinh(\gamma_{_{N+l,k}}) \sin
                        \left(\frac{\pi (N+i) k}{N+M+1} \right)
                        \sin\left(\frac{\pi l i }{M+1}\right),\label{Bogol2}
         \eea
         \end{widetext}
	for $l:N+1\rightarrow N+M$. The transformations \ref{Q2PQ} can similarly be inverted to obtain the inverse Bogoliubov coefficients 
	$\tilde{\alpha}_{lk}, \tilde{\beta}_{lk}$, which 
	by the property of real, symmetric, orthogonal transformation between real space and normal mode, are easy to compute.
	
	\section{Observables}
	\label{App:C}
	
	To study equilibrium relaxation in this system we consider some
        observables to probe the system. In this case, we  use the
        occupation number of the modes of the disjoint system, i. e.,
	\begin{equation}
		\label{oper}
		\hat{n}_m = \hat{a}^\dagger_m \hat{a}_m,\ m=1,2\ldots,(N+M)
	\end{equation}
	Since these operators do not commute with the Hamiltonian ($\hat{H}$)
        after the quench, the time evolution is expected to be
        non-trivial. Using the Bogoliubov transformations and the
        evolution properties of $\hat{c}^\dagger$ and $\hat{c}$ operators, the
        time evolution of the occupation numbers is given by,
        \begin{widetext}
        {\small 
	\begin{equation}
		\label{operevolve1}
		\begin{aligned}
			\langle\hat{n}_m(t)\rangle =&
                        \frac{4}{(N+M+1)(N+1)}
                        \sum_{i,j=1}^{N} \sum_{k,l=1}^{N+M} 
                        \bigg[\cosh\gamma_{_{m,k}} \, \cosh\gamma_{_{m,l}} e^{\frac{i
                              (\omega'_l - \omega'_k)
                              t}{\hbar}}\langle\hat{c}^\dagger_l
                          \hat{c}_k\rangle +\sinh \gamma_{_{m,k}} \cosh \gamma_{_{m,l}} e^{\frac{i
                              (\omega'_l + \omega'_k)
                              t}{\hbar}}\langle\hat{c}^\dagger_l
                          \hat{c}^\dagger_k\rangle \\ &
                          +\cosh \gamma_{_{m,k}} \sinh\gamma_{_{m,l}} \, e^{\frac{i
                              (-\omega'_l - \omega'_k)
                              t}{\hbar}}\langle\hat{c}_l
                          \hat{c}_k\rangle \\ & + \sinh \gamma_{_{m,k}} \sinh\gamma_{_{m,l}} e^{\frac{i
                              (-\omega'_l -\omega'_k)
                              t}{\hbar}}\langle\hat{c}_l
                          \hat{c}^\dagger_k\rangle\bigg]
                        \sin(\frac{\pi j l}{N+M+1})\sin(\frac{\pi j
                          m}{N+1}) \sin(\frac{\pi i
                          k}{N+M+1}) \sin(\frac{\pi i m}{N+1})
		\end{aligned}
	\end{equation}
	}
	for $m:1\rightarrow N$, and
	
	\begin{equation}
		\label{operevolve2}
		\begin{aligned}
			\langle\hat{n}_{N+m}(t)\rangle =&
                        \frac{4}{(N+M+1)(M+1)}
                        \sum_{i=1}^{M}\sum_{j=1}^{M}\sum_{k=1}^{N+M}\sum_{l=1}^{N+M}
                        \bigg[\cosh(\gamma_{_{N+m,k}})\cosh(\gamma_{_{N+m,l}})e^{\frac{i
                              (\omega'_l - \omega'_k)
                              t}{\hbar}}\langle\hat{c}^\dagger_l
                          \hat{c}_k\rangle \\ & + \sinh(\gamma_{_{N+m,k}})\cosh(\gamma_{_{N+m,l}})e^{\frac{i
                              (\omega'_l + \omega'_k)
                              t}{\hbar}}\langle\hat{c}^\dagger_l
                          \hat{c}^\dagger_k\rangle + \cosh(\gamma_{_{N+m,k}})\sinh(\gamma_{_{N+m,l}})e^{\frac{i
                              (-\omega'_l - \omega'_k)
                              t}{\hbar}}\langle\hat{c}_l
                          \hat{c}_k\rangle\\ &+\sinh(\gamma_{_{N+m,k}})\sinh(\gamma_{_{N+m,l}})e^{\frac{i
                              (-\omega'_l -\omega'_k)
                              t}{\hbar}}\langle\hat{c}_l
                          \hat{c}^\dagger_k\rangle\bigg]\times
                        \sin(\frac{\pi (N+j)
                          l}{N+M+1} \times\sin(\frac{\pi j
                          m}{M+1}) \sin(\frac{\pi (N+i) k}{N+M+1})
                        \sin(\frac{\pi i m}{M+1})
		\end{aligned}
	\end{equation}
	\end{widetext}
	for $m:1\rightarrow M$. In order to evaluate the quantum expectation of the number
        operator at any instant of time $t$, we need to obtain the
        correlation matrices
        $\langle\hat{c}^\dagger_i\hat{c}^\dagger_j\rangle$,$\langle\hat{c}_i\hat{c}^\dagger_j\rangle$,
        $\langle\hat{c}^\dagger_i\hat{c}_j\rangle$ and
        $\langle\hat{c}_i\hat{c}_j\rangle$.
	
	Once we obtain the average occupation numbers, it possible to
        obtain the evolution of the energies of the disjoint modes as,
        $\langle\hat{H}_N(t)\rangle=E_N(t) = \sum_{i=1}^{N}
        (\hat{n}_i(t) + 1/2)\hbar \omega_i$ and
        $\langle\hat{H}_M(t)\rangle=E_M(t)=\sum_{i=1}^{M}
        (\hat{n}_{N+i}(t) + 1/2)\hbar \omega_{N+i}$ using the
        occupation numbers we have computed above.
	
	\section{Long time averages}
	\label{App:D}
	
	From \ref{operevolve1} and \ref{operevolve2}, it is
        easy to see that that the long time average will include only
        the diagonal elements as only for the those values will the
        exponential factor vanish.
	
	Hence, the long time average of the expectation value of the
        number operator is given by
	
	\begin{widetext}
	\begin{equation}
		\label{operav1}
		\begin{aligned}
			\overline{\langle\hat{n}_m(t)}\rangle
                        =&\frac{4}{(N+M+1)(N+1)}
                        \sum_{i=1}^{N}\sum_{j=1}^{N}\sum_{k=1}^{N+M}
                        \bigg[\cosh^2(\gamma_{_{mk}})\langle\hat{c}^\dagger_k
                          \hat{c}_k\rangle +\sinh^2(\gamma_{_{mk}})
                          \langle\hat{c}_k
                          \hat{c}^\dagger_k\rangle\bigg]\\ & \times
                        \sin(\frac{\pi j k}{N+M+1})\sin(\frac{\pi j
                          m}{N+1}) \sin(\frac{\pi i k}{N+M+1})
                        \sin(\frac{\pi i m}{N+1})
		\end{aligned}
	\end{equation}
	for $m:1\rightarrow N$, and
	\begin{equation}
		\label{operav2}
		\begin{aligned}
			\overline{\langle\hat{n}_{N+m}(t)}\rangle =&
                        \frac{4}{(N+M+1)(M+1)}
                        \sum_{i=1}^{M}\sum_{j=1}^{M}\sum_{k=1}^{N+M}
                        \bigg[\cosh^2(\gamma_{_{N+m,k}})\langle\hat{c}^\dagger_k
                          \hat{c}_k\rangle+\sinh^2(\gamma_{_{N+m,k}})\langle\hat{c}_k
                          \hat{c}^\dagger_k\rangle\bigg]\\ &\times
                        \sin(\frac{\pi (N+j) k}{N+M+1})\sin(\frac{\pi
                          j m}{M+1}) \sin(\frac{\pi (N+i) k}{N+M+1})
                        \sin(\frac{\pi i m}{M+1})
		\end{aligned}
	\end{equation}
	for $m:1\rightarrow M$.
	\end{widetext}

\section{\label{App:F}Covariance Matrix}

For Gaussian states, the information is encoded in the second moment
which is also encoded in the covariance matrix in the phase space.
Structurally, the covariance matrix is given as 
\bea \sigma_{NM} =\left(
\begin{array}{ll}
\sigma_{xx} &  \sigma_{xp} \\
\sigma_{xp} &  \sigma_{pp}
\end{array}
\right) \eea The covariance matrix before the quench, in the
normal modes of the `non-quenched' Hamiltonian, is given as
 \bea
\sigma_{NM} =\left(
\begin{array}{llllll}
a_{11} & ..& a_{1n} &  b_{11} & ..& b_{1n} \\
  ..   & ..&   ..   &   ..    & ..& ..\\
a_{n1} & ..& a_{nn} &  b_{n1} & ..& b_{nn} \\
b_{11} & ..& b_{1n} & \tilde{a}_{11} & .. & \tilde{a}_{1n}\\
  ..   & ..&   ..   &   ..    & ..& ..\\
b_{n1} & ..& b_{nn} & \tilde{a}_{n1} & .. & \tilde{a}_{nn}  
\end{array}
\right) \label{InCov}
\eea
If we go to the phase space corresponding to the configuration space variables of the system, the covariance matrix becomes
\bea 
\sigma = S^{T} \sigma_{NM} S, \label{Config1} 
\eea where $S$ is given by 
 \bea S =\left(
\begin{array}{llll}
S_N & 0& 0& 0\\
0 & S_M& 0& 0\\
0 & 0&S_N & 0\\
0 & 0& 0& S_M\\
\end{array}
\right)
\eea
which is a similarity transformation from the configuration space to the
normal modes and 
\bea
S_K =\sqrt{\frac{2}{K+1}}\left(
\begin{array}{lll}
\sin{\frac{\pi}{K+1}} & ..& \sin{\frac{K\pi}{K+1}}\\
.. & ..& .. \\
\sin{\frac{K\pi}{K+1}} & ..& \sin{\frac{K^2\pi}{K+1}}\\
\end{array}
\right).  \eea
$K$ takes values $N, M$. Clearly, being a symmetric orthogonal matrix $ S$ is
its inverse as well. Since it is easier to go to the normal modes of quenched system from the configuration space we cast the  covariance matrix
in the configuration space through \ref{Config1}. 

\emph{Post quenching}, the system can
be expressed in terms of normal modes of the quenched system and we
cast the covariance matrix in the new normal mode basis of the quenched system to study its
time evolution. The recasting in the normal mode basis is done again
by an orthogonal transformation
\begin{widetext}
{\small
\bea
\tilde{S}_{N+M} =\sqrt{\frac{2}{N+M+1}}\left(
\begin{array}{llllll}
\sin{\frac{\pi}{N+M+1}} & ..& \sin{\frac{(N+M)\pi}{N+M+1}} &0&..&0\\
.. & ..& .. &..&..&..\\
\sin{\frac{(N+M)\pi}{N+M+1}} & ..& \sin{\frac{(N+M)^2\pi}{N+M+1}}&0&..&0\\
0&..&0&\sin{\frac{\pi}{N+M+1}} & ..& \sin{\frac{(N+M)\pi}{N+M+1}}\\
.. & ..& .. &..&..&..\\
0&..&0&\sin{\frac{(N+M)\pi}{N+M+1}} & ..& \sin{\frac{(N+M)^2\pi}{N+M+1}}
\end{array}
\right).
\eea
}
\end{widetext}
Therefore, the covariance matrix post quenching is given as 
\bea \sigma_{PQ} = \tilde{S}_{N+M}(S^{T} \sigma_{NM} S)
\tilde{S}_{N+M}^{T}.  \label{PQS}
\eea 
Let us call
\bea
\sigma_{PQ} =\left(
\begin{array}{llllll}
a'_{11} & ..& a'_{1n} &  b'_{11} & ..& b'_{1n} \\
  ..   & ..&   ..   &   ..    & ..& ..\\
a'_{n1} & ..& a'_{nn} &  b'_{n1} & ..& b'_{nn} \\
b'_{11} & ..& b'_{1n} & \tilde{a}'_{11} & .. & \tilde{a}'_{1n}\\
  ..   & ..&   ..   &   ..    & ..& ..\\
b'_{n1} & ..& b'_{nn} & \tilde{a}'_{n1} & .. & \tilde{a}'_{nn}  
\end{array}
\right). \label{InCov2}
\eea
Further, in the normal modes basis, we have
$N+M$ decoupled harmonic oscillators whose time evolution is obtained
as a transformation given by matrix
%
{\small
\bea
U_{t} =\left(
\begin{array}{llllll}
\cos{\omega_1 t} &..& 0 & \frac{\sin{\omega_1 t}}{\omega_1} &..&0\\
.. &.. &..& .. &..&..\\
0 & .. & \cos{\omega_{_{N+M}} t} &0 &..& \frac{\sin{\omega_{_{N+M}} t}}{\omega_{_{N+M}}}\\
-\omega_1\sin{\omega_1 t} &..& 0 & \cos{\omega_1 t} &..&0\\
.. &.. &..& .. &..&..\\
0 & .. & -\omega_{_{N+M}} \sin{\omega_{_{N+M}} t} &0 &..& \cos{\omega_{_{N+M}} t}\\
\end{array}
\right),
\eea
}
%
It is important to note that this is not unitary in the phase space. Thus the time evolution
drives the covariance matrix to \bea \sigma_{PQ} (t) = U_t^{T}
\sigma_{PQ} U_t = \chi ^T \sigma_{NM} \chi,  \eea 
with \bea \chi = S \tilde{S}_{N+M}^{T}U_t.  \eea
The generic form of the off-diagonal terms in the covariance matrix post quench are given as
\begin{widetext}
\bea
(\sigma_{PQ;xx})_{ij} &=& a'_{ij} \cos{\omega_j t}\cos{\omega_i t} - b'_{ij}\omega_j\sin{\omega_j t}\cos{\omega_i t}-b'_{ij}\omega_i\cos{\omega_j t}\sin{\omega_i t} 
+ \tilde{a}'_{ij}\omega_i\omega_j\sin{\omega_j t}\sin{\omega_i t}\nonumber\\
\label{eq:CovFin}
(\sigma_{PQ;xp})_{ij} &=& \omega_j^{-1} a'_{ij} \sin{\omega_j t}\cos{\omega_i t} + b'_{ij}\cos{\omega_i t}\cos{\omega_j t}- \frac{\omega_i}{\omega_j}b'_{ij}\sin{\omega_i t}\sin{\omega_j t} -\tilde{a}'_{ij}\omega_i\sin{\omega_i t}\cos{\omega_j t}.
\eea
\end{widetext}
For the vacuum state, evidently $a_{ij}|_{i\neq  j}=0, b_{ij}=0 \Rightarrow b'_{ij}=0$.  Therefore, the off-diagonal elements of the quenched covariance matrix become purely oscillatory in nature.
The long-time averages over such covariance matrices can be effectively replaced by a diagonal Gaussian covariance matrix, which acquire GGE form. 
The diagonal elements, obtained from \ref{PQS} and \ref{InCov2}, give the occupancy across different modes, which is exactly the first term on the RHS of \ref{Emission}.
Similar exercise for any pre-quenching initial eigenstate reproduces \ref{Emission}.

\section{\label{APP:G}Fate of a general Fock basis state under quench}
As discussed previously, any non-vacuum eigenstates will bring corrections as depicted in \ref{Emission}. If the initial state is taken to be an eigenstate of the pre-quenched Hamiltonian, i.e.
\bea
|\Psi_{in}\rangle = |n_1,n_2,...\rangle_M\otimes|n_{N+1},n_{N+2},..\rangle_N \label{Excited}
\eea
the off diagonal parts of the initial covariance matrix are given as
 (by writing the individual $x_i$ and $p_j$'s in terms of the
creation and annihilation operators)
\bea
\sigma_{x_i x_j}&=& \langle \Psi | \frac{\hat{x}_i \hat{x}_j + \hat{x}_j \hat{x}_i}{2} |\Psi \rangle \big|_{i\neq j} =0,\\ 
\sigma_{x_i p_j}&=& \langle \Psi | \frac{\hat{x}_i \hat{p}_j + \hat{p}_j \hat{x}_i}{2} |\Psi \rangle =0. \label{OffD_InCov}
\eea
owing to $b_{ij} =0$ in \ref{eq:CovFin} for the energy eigenstates \ref{Excited}. 
We note the following points: The initial states, for which  $b_{ij} =0$, the long time average of the covariance matrix will lead to {\it thermal} form. 
Clearly these states satisfy the necessary criterion of a diagonal covariance matrix. It is also evident from the oscillatory behaviour of the $b'_{ij}$ that for any initial finite $b_{ij}$ the late time $b'_{ij} \rightarrow 0$. However, for the initial states, which have $b_{ij} \nless \infty$ may potentially violate {\it thermalization.} 
A simple example of such states in the Hilbert space may be the class of separable states of the kind
\bea
|\Psi\rangle = \sum_{n_1,n_2,..,n_M,..n_{N+M}} \!\!\!\!\!\!\!\!\!\!\!\!\!\!\! c_{n_1}....c_{n_{N+M}} |n_1 ..n_M,...n_{N+M}\rangle,
\eea
where for at least one $c_{n_m},$ we should have
$|c_{n_m}|^2 \sim \frac{1}{(n_m)^{1/2-\epsilon}}$, at least for $n_m \gg 1$. This condition is sufficient to ensure the normalization of the state, however, for such states
$\langle x_m \rangle =\sum_{n_m} c_{n_m} (\sqrt{n_m}c^*_{n_m-1}+\sqrt{n_m+1}c^*_{n_m+1}) \rightarrow \infty$. Therefore, the late time $b'_{ij}$ in the  covariance matrix can still survive, forbidding the system to land up in a GGE configuration. {\it  Such states may well be very delocalized as we just require one of the $c_{n_m}$ to have the above mentioned property. Therefore a fine distribution of over various energy eigenstates is permissible rendering localization inefficient. These states have the property that some of the conserved charges are divergent, for which already failure of GGE has been argued \cite{Kormos}.}  Such states are highly energetic ones, but definitely physical, as they belong to square integrable class. The analysis of  measure of such states in the Hilbert space is an interesting topic, which we will pursue in a subsequent work. These kind of states may also be relevant in the case of black hole physics, where non-thermality of the configuration may be related to memory of the 
in-
state, while the energetics of such states may shed some light on the {\it firewall proposal} \cite{Almheiri}.

Coming back to the eigenstates, the comparison  of the diagonal elements of the time averaged covariance matrix expectedly recovers \ref{opergibbs1},\ref{opergibbs2}, which 
supports the GGE description. We now further probe the excited eigenstates with their long time average expectation values.

Clearly, in the free harmonic oscillator system, all observables are described by the correlators $\langle \hat{c}_k (t)\hat{c}^{\dagger}_{k'}(t) \rangle$. 
As discussed in the previous appendices, the long time average for any observable will be solely determined by $\langle \hat{c}_k (0)\hat{c}^{\dagger}_{k}(0) \rangle$. 
Therefore, we can use \ref{Emission} for determining such quantities per unit lattice. Clearly, for initial excited states, the Gaussian (GGE) description will be corrected. 
The deviation from such a description is characterized by 
\bea
\Delta_G =\frac{\sum_j\left(|\tilde{\alpha}_{k j}|^2 n_j+ |\tilde{\beta}_{k j}|^2
        n_j\right)/N}{\sum_l |\tilde{\beta}_{kl}|^2/N}
\eea
In the limit $N \rightarrow \infty$, 
\bea 
\Delta_G  &\rightarrow& {\cal O}\left( \frac{\sum_j n_j}{N}  \right).
\eea
Therefore, in the thermodynamic limit, for any finitely excited state (sub-part of the full system) the effective description is essentially GGE.
Also, we note that quantum states for which  $\Delta_G \sim {\cal O}\left(1\right)$ will not relax to a generalized thermal setting effectively.

\section{\label{APP:H} Correlation between the lattice sites}
In order to check the effectiveness of GGE relaxation, we explore the correlation between different lattice sites, post quenching. Therefore, the operator to explore would be 
\bea
\langle \psi| \hat{n}_{k_1 }'\hat{n}_{k_2}'|\psi \rangle  = {}_M\langle 0|{}_N\langle 0|\hat{c}_{k_1 }^{\dagger}\hat{c}_{k_1 }\hat{c}_{k_2 }^{\dagger}\hat{c}_{k_2 }|0\rangle_M|0\rangle_N,
\eea
which measures the correlation between $k_1-$th and $k_2-$th site.
Writing 
\bea
\hat{c}_{k}=\sum_{k'}\tilde{\alpha}_{kk'}\hat{a}_{k'}+\tilde{\beta}_{kk'}\hat{a}_{k}^{\dagger},
\eea
we obtain,
\begin{widetext}
{\small
\begin{equation}
\langle \psi| \hat{n}_{k_1 }'\hat{n}_{k_2}'|\psi \rangle  = \sum_{lm} \left( \tilde{\beta}_{k_1l}\tilde{\beta}_{k_2m}^*\tilde{\alpha}_{k_1m}\tilde{\alpha}_{k_2l}^* + \tilde{\beta}_{k_1l}\tilde{\beta}_{k_2l}^*\tilde{\alpha}_{k_1m}\tilde{\alpha}_{k_2m}^* + |\tilde{\beta}_{k_1l}\tilde{\beta}_{k_2m}|^2 \right)\, . \label{Crosscorrelation}
\end{equation} 
}
\end{widetext}
Owing to the real character of the Bogoliubov coefficients along with \ref{Bogol1}, \ref{Bogol2} and the relation (for the Bogoliubov coefficients to yield the correct commutation relations)
\bea
\sum_{k'} \tilde{\alpha}_{kk'}\tilde{\beta}_{lk'}-\tilde{\alpha}_{lk'}\tilde{\beta}_{kk'} = 0,
\eea
we obtain in that the first two terms in the RHS of \ref{Crosscorrelation} are measure to the ($Q_{k_1}Q_{k_2}$,  $Q_{k_1}P_{k_2}$, etc.) correlations post quench, which vanish. Therefore, only the third term survives and we have
\bea
\langle \psi| \hat{n}_{k_1 }'\hat{n}_{k_2}'|\psi \rangle  =\sum_{lm} |\tilde{\beta}_{k_1l}\tilde{\beta}_{k_2m}|^2 = \langle n_{k_1}'\rangle \langle n_{k_2}'\rangle,
\label{correlation} 
\eea
which is obtained for the initial vacuum state pre quenching, i.e. $n_i =0, \hspace{0.1 cm} \forall i$ in \ref{Emission}.
Therefore, post quench the system lands up in a configuration in which there is no real quantum correlation between different lattice sites, as suited for a Gibbs Generalized ensemble.
If the density matrix description of this state post quench is indeed that of a Gibbs Generalized Ensemble, then the two point correlation can be obtained as
\bea
\langle\hat{n}_{k_1 }'\hat{n}_{k_2}'\rangle = \frac{Tr[n_{k_1}'n_{k_2}' \rho_{GGE}]}{Tr[\rho_{GGE}]}.
\eea
Since $n_K$s are  conserved charges of the system post quench,
\bea
\rho_{GGE} = e^{-\sum_k \lambda_k n_k'}.
\eea
A simple computation yields,
\bea
\langle\hat{n}_{k_1 }'\hat{n}_{k_2}'\rangle =\frac{e^{-\lambda_{k_1}}}{(1-e^{-\lambda_{k_1}})}\frac{e^{-\lambda_{k_2}}}{(1-e^{-\lambda_{k_2}})}.
\eea
Since previously we obtained,
\bea 
e^{\lambda_k} =1 + \frac{1}{\langle n_k(0)'\rangle},
\eea
we finally have,
\bea
\langle\hat{n}_{k_1 }'\hat{n}_{k_2}'\rangle = \langle n_{k_1}'\rangle \langle n_{k_2}'\rangle,
\eea
relating the two-point correlation post quench to that of classical correlation i.e. 
$$ \langle n_{k_1}' n_{k_2}'\rangle-\langle n_{k_1}'\rangle\langle n_{k_2}'\rangle=0,$$ 
as in \ref{correlation}.

\bibliography{Reference.bib}
\end{document}